\documentclass[twocolumn,superscriptaddress,preprintnumbers]{revtex4-1}

\usepackage{graphicx}
\usepackage[caption=false]{subfig}
\usepackage{float}
\usepackage{epstopdf}
\usepackage{dcolumn}
\usepackage{color}
\usepackage{soul}
\setstcolor{red}
\usepackage{natbib}
\usepackage{amsmath}

\newcommand{\nio}{Na$_2$IrO$_3$}
\newcommand{\lio}{Li$_2$IrO$_3$}

\newcommand{\rucl}{$\alpha$-RuCl$_3$}

\definecolor{britishracinggreen}{rgb}{0.0, 0.26, 0.15}
\definecolor{deeplilac}{rgb}{0.6, 0.33, 0.73}

\definecolor{schrift}{cmyk}{.4,1,1,0}

\newcommand{\SD}{$S_{\text{d}}$}

\begin{document}

\title{
Signatures of low-energy fractionalized excitations in {\rucl} \\ from field-dependent microwave absorption }

\author{C. Wellm}
{\thanks{These authors contributed equally to this work.}
\affiliation{Leibniz Institute for Solid State and Materials Research IFW Dresden,
01171 Dresden, Germany}
\affiliation{Institut f\"ur Festk\"orper- und Materialphysik, Technische Universit\"at Dresden, 01062 Dresden, Germany}
\author{J. Zeisner}
\thanks{These authors contributed equally to this work.}
\affiliation{Leibniz Institute for Solid State and Materials Research IFW Dresden,
01171 Dresden, Germany}
\affiliation{Institut f\"ur Festk\"orper- und Materialphysik, Technische Universit\"at Dresden, 01062 Dresden, Germany}
\author{A. Alfonsov}
\affiliation{Leibniz Institute for Solid State and Materials Research IFW Dresden, 01171 Dresden, Germany}
\author{A. U. B. Wolter}
\affiliation{Leibniz Institute for Solid State and Materials Research IFW  Dresden, 01171 Dresden, Germany}
\author{M. Roslova}
\affiliation{Fakult\"{a}t Chemie und Lebensmittelchemie, Technische Universit\"at Dresden, 01062 Dresden, Germany}
\author{A. Isaeva}
\affiliation{Fakult\"{a}t Chemie und Lebensmittelchemie, Technische Universit\"at Dresden, 01062 Dresden, Germany}
\author{T. Doert}
\affiliation{Fakult\"{a}t Chemie und Lebensmittelchemie, Technische Universit\"at Dresden, 01062 Dresden, Germany}
\author{M. Vojta}
\affiliation{Institut f\"ur Theoretische Physik, Technische Universit\"at Dresden, 01062 Dresden, Germany}
\author{B. B\"{u}chner}
\affiliation{Leibniz Institute for Solid State and Materials Research IFW Dresden, 01171 Dresden, Germany} \affiliation{Institut f\"ur Festk\"orper-
und Materialphysik, Technische Universit\"at Dresden, 01062 Dresden, Germany}
\author{V. Kataev}
\affiliation{Leibniz Institute for Solid State and Materials Research IFW Dresden, 01171 Dresden, Germany}



\begin{abstract}

Topologically ordered states of matter are generically characterized by excitations with quantum number fractionalization. A prime example is the
spin liquid realized in Kitaev's honeycomb-lattice compass model where spin-flip excitations fractionalize into Majorana fermions and Ising gauge
fluxes. While numerous compounds have been proposed to be proximate to such a spin-liquid phase, clear-cut evidence for fractionalized excitations is
lacking.
Here we employ microwave absorption measurements to study the low-energy excitations in {\rucl} over a wide range of frequencies, magnetic fields,
and temperatures, covering in particular the vicinity of the field-driven quantum phase transition where long-range magnetic order disappears. In
addition to conventional gapped magnon modes we find a highly unusual broad continuum characteristic of fractionalization which -- most remarkably -
extends to energies below the lowest sharp mode and to temperatures significantly higher than the ordering temperature, and develops a  gap of a nontrivial origin in strong magnetic fields. Our results unravel the
signatures of fractionalized excitations in {\rucl} and pave the way to a more complete understanding of the Kitaev spin liquid and its
instabilities.

\end{abstract}

\maketitle

\section{Introduction \label{intro}}

Spin liquids -- low-temperature states of local-moment insulators devoid of symmetry-breaking long-range order -- constitute a class of most
fascinating states of matter. They are characterized by topological order and fractionalization, i.e., local excitations decay into fractionalized
constituents, which typically leads to continua instead of sharp modes in the dynamic response of the material.
The seminal work of Kitaev  \cite{Kitaev2006} has introduced a particular spin-liquid model, with compass interactions on the honeycomb lattice,
whose fractionalized excitations are dispersive Majorana fermions and static Ising gauge fluxes (visons). Subsequently,
it has been proposed that this model may be approximately realized in certain Mott insulators with strong spin-orbit coupling
\cite{Jackeli2009,Chaloupka2010}.
Candidate materials \cite{Winter2017c} are \nio, different polytypes of \lio, and \rucl. While these materials display long-range magnetic order
below a small N\'eel temperature $T_{\rm N}$, likely due to the presence of additional interactions, it is believed that they are proximate to a
Kitaev spin-liquid phase. As a result, signatures of Kitaev physics are expected in various physical probes including the excitation spectrum, and
pressure or magnetic field might even stabilize a spin-liquid ground state.

In {\rucl} \cite{Plumb2014,Sears2015}, currently considered the most promising Kitaev material, unconventional behavior has been reported in a number
of experimental probes. Antiferromagnetic order which occurs in single crystalline samples of \rucl\  without stacking faults at $T_{\rm N}$ in a range 7\,-8\,K can be completely suppressed by application of a moderate  in-plane magnetic field $\mu_0H_{\rm c}\sim 7$\,T (see, e.g., \cite{Sears2015,Sears2017,Banerjee2017b,Wolter2017,Hentrich2017,Kasahara2018}). Remarkably, recent inelastic neutron scattering (INS) experiments \cite{Banerjee2017a} have revealed an unusually broad magnetic response near the Brillouin-zone center, $\vec{q} = 0$, which persists up to temperatures of 100\,K, the energy scale of the estimated
Kitaev coupling in \rucl\ \cite{Sandilands2015,Nasu2016,Banerjee2016}. This elevated-energy response apparently unrelated to magnetic order may thus be
consistent \cite{Banerjee2017a} with continuum scattering off Majorana excitations inherent in the Kitaev spin liquid \cite{Knolle2014a,Balents2016}. The emergence of the Majorana fermions in \rucl\ was further suggested by a combined specific heat and INS study in Ref.~\cite{Do2017}.
In addition, sharper spin-wave-like excitations at $\vec{q} = 0$ with an apparent gap $\Delta_{\rm sw}^{\vec{q}=0}\sim 2.7$\,meV at zero magnetic
field have been found below $T_{\rm N}$. The existence of these magnon modes has been confirmed by other experimental techniques
\cite{Little2017,Wang2017,Ponomaryov2017} and the low-temperature study of their field dependence has revealed that $\Delta_{\rm sw}^{\vec{q}=0}(H)$
still has a sizable minimum value of $\sim 1$\,meV at the critical field  $\mu_0H_{\rm c}\sim 7$\,T
\cite{Little2017,Wang2017,Ponomaryov2017}.

To clarify the nature of the excitation spectrum of {\rucl} we have measured the microwave absorption (MWA) of high-quality single crystals in the
temperature range $3 - 30$\,K as function of the magnetic field up to $16$\,T, employing a high-field/high-frequency electron spin resonance (ESR)
setup where the signal loss is proportional to the imaginary part of the dynamic spin susceptibility $\chi^{\prime\prime}(\omega)$ at the chosen
excitation frequency $\nu\!=\!\omega/(2\pi)$ and wavevector $\vec{q} = 0$. Working with fixed selected
frequencies, we covered the range of $\nu = 70 - 660$\,GHz ($0.3 - 2.8$\,meV). We note that energies below $1.5$\,meV were inaccessible in the
INS study in Ref.~\cite{Banerjee2017a} (1.0\,meV in Ref.~\cite{Banerjee2017b}) due to the presence of the elastic peak at $\vec{q} = 0$.
Also, to our knowledge, the magnetic field dependence of the dynamic response $\chi^{\prime\prime}(\omega)$  at $T>T_{\rm N}$ in this important energy range was not addressed so far (cf. Refs.~\cite{Little2017,Wang2017,Ponomaryov2017,Shi18,Reschke18,Wu18}). Besides the sharp resonant modes at $T<T_{\rm N}$ reported earlier \cite{Little2017,Wang2017,Ponomaryov2017}, we observe non-resonant, magnetic field dependent absorption, which is particularly strong for energies between $100$ and $360$\,GHz
($0.4 - 1.5$\,meV), i.e., \emph{below} the apparent magnon gap, which persists up to temperatures significantly larger than $T_{\rm N}$, and develops a gap of a nontrivial origin in magnetic fields exceeding 7\,T. This novel finding
provides  evidence for an excitation continuum extending down to low energies of the order
$\sim 0.4$\,meV, likely arising from fractionalization.\\


\section{Experimental details}

\subsection{Crystal synthesis and characterization}\label{samples}

%
Single crystals used in this study have been grown at TU Dresden. For the synthesis, pure ruthenium-metal powder (99.98 \%, Alfa Aesar) was filled
into a quartz ampoule under argon atmosphere, together with a sealed silica capillary containing chlorine gas (99.5 \%, Riedel-de Ha\"{e}n). The
chlorine gas was dried prior to use by passing through concentrated H$_2$SO$_4$ and CaCl$_2$, the ruthenium powder used was without further
purification. The molar ratio of the starting materials was chosen to ensure in-situ formation of RuCl$_3$ and its consequent chemical transport
according to the reaction: RuCl$_3$(s) + Cl$_2$ = RuCl$_4$(g) \cite{Binnewies2012}. After sealing the reaction ampoule under vacuum, the
chlorine-containing capillary inside was broken in order to release the gas. The ampoule was kept in the temperature gradient between 750 $^\circ$C
(source) and 650 $^\circ$C (sink) for 5 days. The obtained crystalline product represented pure {\rucl} (according to powder X-ray diffraction)
without inclusions of ruthenium. Single-crystal X-ray diffraction studies (Apex II diffractometer, Bruker-AXS, Mo K$\alpha$ - radiation) and EDXS
(Hitachi SU 8020 SEM, 20 kV with an Oxford Silicon drift detector XMax$^{\rm N}$) of the crystals have confirmed a monoclinic structure
\cite{Johnson2015} and the nominal composition. The crystals are black with a shiny surface, of a size of several millimeters along the
\emph{ab}-plane, and of a thickness about 0.2\,mm [Fig.~S3, right panel, in Supplemental Material (SM) \cite{Suppl}]. MWA measurements were performed on three samples from the same batch. They all show a similar AFM
ordering temperature $T_{\rm N} = 8$\,K which has been determined through magnetization measurements with a vibrating sample magnetometer
(Quantum Design) with superconducting quantum interference device detection (SQUID-VSM).

\subsection{Microwave absorption measurements}\label{mwa}

%
For measurements of MWA a home made high-frequency/high-field ESR (HF-ESR) spectrometer was employed. A vector network analyzer (PNA-X from Keysight
Technologies) was used for generation and detection of microwaves in the frequency range from 70\,GHz to 330\,GHz. In addition, an
amplifier/multiplier chain (AMC from Virginia Diodes Inc.) allowed for measurements with frequencies up to 480\,GHz, detected using a hot electron
InSb bolometer (QMC Instruments). The probe-head with the sample was mounted in the variable temperature insert of the superconducting magnet system
(Oxford Instruments) enabling field sweeps up to 16\,T. Measurements up to 660\,GHz were performed using a vector network analyser from ABmillimetre
and a 14 T magnet from Cryogenic Limited. Measurements were carried out in transmission mode using a Faraday configuration, i.e. with the $k$-vector
of the microwaves oriented parallel to the direction of the external magnetic field, which in turn was parallel to the sample \emph{ab}-plane. MWA
spectra at each temperature were recorded at fixed selected frequencies (see SM \cite{Suppl} and related Refs.~\cite{Eilers94,Kullmann84,Brunel02,Sosin08} for further technical details of the MWA measurements).

\section{Results \label{results}}

\begin{figure*}[t]
	\centering
	\includegraphics[clip,width=1.5\columnwidth]{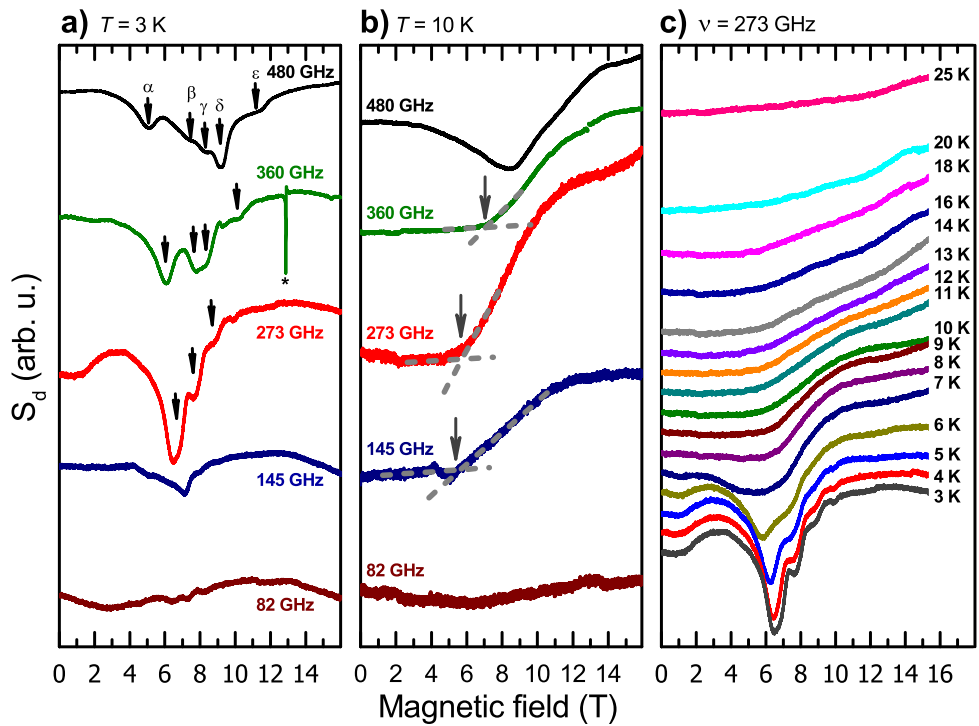}
	\caption{ {\bf Microwave absorption spectra.}
		(a) Signal at the detector \SD\ measured at $T = 3$\,K and various frequencies of the microwave radiation $\nu$. Spectra are shifted vertically for
		clarity. Peaks "$\alpha, \beta, \gamma, \delta, \epsilon $" are frequency dependent resonant magnon-like modes in agreement with
		Refs.~\cite{Wang2017,Little2017,Ponomaryov2017} [see also Fig.~S4(a) in SM \cite{Suppl}]. Narrow signal (*) at $\nu = 360$\,GHz is from a reference sample (DPPH). 
		(b) Set of \SD$(H)$ spectra at $T = 10$\,K for various $\nu$. Arrows mark the inflection point. Dashed lines are guides for the eye.
		The DPPH signal at $\nu = 360$\,GHz is not shown here (cf. Fig.~S2 in SM \cite{Suppl}). 
		(c) Temperature dependence of \SD\ at $\nu = 273$\,GHz.
		Data shown are obtained from as-recorded spectra by subtracting the featureless flat spectrum measured at 30\,K from spectra measured at lower
		temperatures. The spectra are shifted vertically for clarity; in (c) the offset is proportional to temperature.} \label{fig:spectra_Fdep_Tdep}
\end{figure*}

Characteristic MWA spectra (signal at the detector \SD($H$) as a function of field at a given constant $\nu$) for $\mathbf{H}\parallel ab$-plane are
summarized in Fig.~\ref{fig:spectra_Fdep_Tdep}. Here, a smaller value of \SD\, corresponds to a larger absorption of microwaves by the sample and
{\it vice versa}.
Fig.~\ref{fig:spectra_Fdep_Tdep}(a) shows the frequency dependence of the MWA signal at 3\,K. At a high frequency of 480\,GHz one observes a set of
lines labeled $\alpha, \beta, \gamma, \delta$ and $\epsilon$ which shift by lowering $\nu$ to 360\,GHz indicating their resonant nature. These lines
are present only at $T<T_{\rm N} = 8$\,K. Their position and the $\nu$-dependence is in agreement with observations interpreted in terms of resonant magnon
modes at $\vec{q} = 0$ reported in Refs.~\cite{Wang2017,Ponomaryov2017} (see also Fig.~S4(a) in SM \cite{Suppl}). Remarkably, by approaching the minimum excitation gap
$ \Delta_{\rm sw}\sim 250$\,GHz for these resonance modes reported in Refs.~\cite{Little2017,Wang2017,Ponomaryov2017}, the spectrum evolves into a pronounced
maximum of MWA centered at $H_{\rm c}\sim 7$\,T with some residual smeared  structure. This response does not shift upon change of frequency within
experimental uncertainties, which suggests a non-resonant nature of the absorption spectrum (see also Figs.~S1 and S4 in SM \cite{Suppl}).

The strong peak in the MWA spectrum in the range $6 - 8$\,T starts to develop at temperatures right below $T_{\rm N} = 8$\,K
[Fig.~\ref{fig:spectra_Fdep_Tdep}(c)]. Above $T_{\rm N}$, the spectra at $\nu \leq 360$\,GHz become flat for magnetic fields below $\sim 6$\,T
showing no field dependence of MWA [Figs.~\ref{fig:spectra_Fdep_Tdep}(b) and (c)]. Only in the vicinity of $T_{\rm N}$ and at $\nu > 360$\,GHz the
MWA signal still exhibits a broad bump due to strongly broadened, rapidly decaying resonance modes $\alpha, \beta, \gamma, \delta$ and $\epsilon$
[cf. top curves in Figs.~\ref{fig:spectra_Fdep_Tdep}(a) and (b)]. A remarkable novel feature of the MWA spectrum at $T > T_{\rm N}$ and $\nu \leq
360$\,GHz is the occurrence of an inflection point in the \SD$(H)$ dependence marked by arrows in Fig.~\ref{fig:spectra_Fdep_Tdep}(b) which shifts
only slightly with changing the frequency. Above this point \SD\ begins to increase indicating a decrease of the MWA by the sample. Importantly, all
non-resonant MWA features are most prominent in the frequency window $\sim 100 - 360$\,GHz. In particular, they rapidly decrease at $\lesssim
150$\,GHz and finally vanish below $\sim 100$\,GHz [see Fig.~\ref{fig:spectra_Fdep_Tdep}(a,b) and the $T - H$ maps of the microwave absorption
plotted for different excitation frequencies in Fig.~S5 in SM \cite{Suppl}].
We note that the major features of the magnetic response discussed above are specific for the $\mathbf{H}\parallel ab$-plane geometry and were not observed for the $\mathbf{H}\parallel c$-axis orientation, similar to, e.g.,  Ref.~\cite{Majumder2015}, where the change of magnetic properties upon variation of the magnetic field in the $c$-axis field geometry became much milder.\\

The contour plots $S_{\rm d}^{\rm norm}(T,H)$ presented in Fig.~\ref{fig:phase_diagram}(a) were obtained from recorded \SD$(H)$ spectra by first
subtracting the \SD\ value at a field of 15.5\,T, and afterwards by subtracting the spectrum measured at 30\,K from the spectra measured at other
temperatures.
To obtain the contour plots $S_{\rm d}^{\rm norm}(\nu,H)$ shown in Fig.~\ref{fig:phase_diagram}(b), the respective value $S_{\rm d}(\nu,15.5 \rm{\, T})$ was subtracted from all the spectra, similar to the procedure described above. Since at different microwave frequencies the sensitivity of the spectrometer may vary, and also since for this kind of measurements several experimental setups were employed, the following normalization of the data was applied for plotting  Fig.~\ref{fig:phase_diagram}(b): The spectra, except the one at  $\nu =
82$\,GHz, were divided by a respective difference value $\Delta S_{\rm d} =
\lvert S_{\rm d}(\nu,15.5 \rm{\, T}) - S_{\rm d}(\nu,0 \rm{\, T})\rvert $. To normalize the flat signal at  $\nu = 82$\,GHz the value $\Delta S_{\rm d}$ from the measurement at a next higher frequency $\nu = 145$\,GHz was used. 
In both figures the gradual change of the color from "cold" violet  to "warmer" colors such as yellow and red corresponds to an increase of the absorption in the	sample.
However, due to a different data treatment the scales in Figs.~\ref{fig:phase_diagram}(a) and (b) are not quantitatively comparable.

\begin{figure*}[!htb]
	\centering
	\includegraphics[width=2\columnwidth]{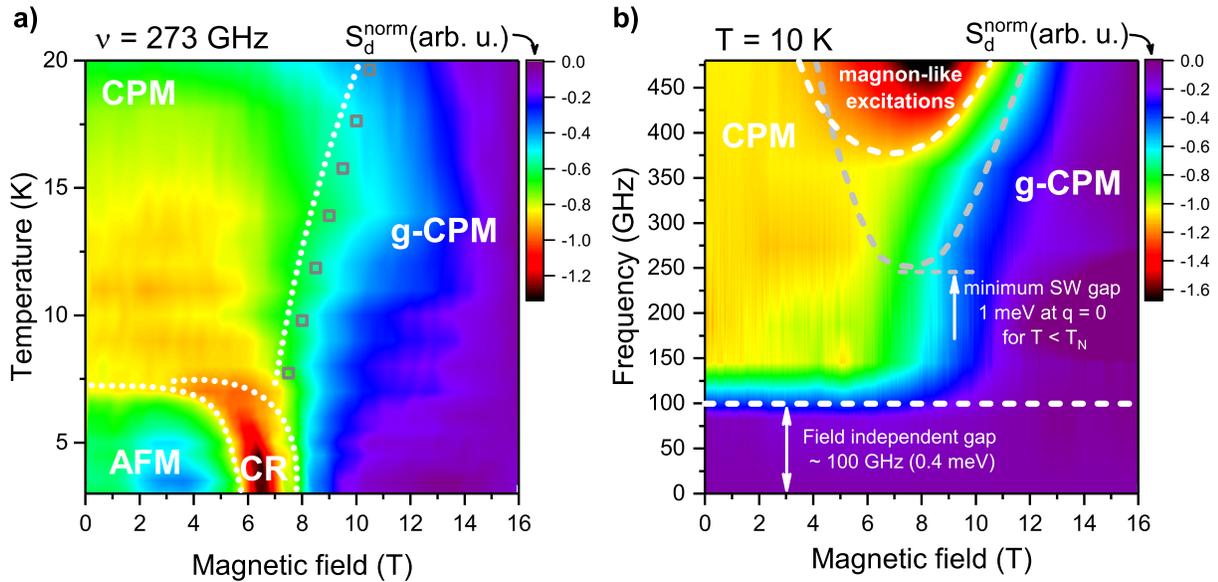}
	\caption{
		{\bf Schematic MWA-based phase diagram.} The diagram is sketched by lines drawn on the color-coded presentations of the MWA measurements at $\nu =
		273$\,GHz  and various temperatures {\bf(a)}, and at $T = 10$\,K and various frequencies {\bf (b)},  indicating major changes in the field dependence
		of the MWA. To plot the maps $S_{\rm d}^{\rm norm}(T,H)$ the recorded \SD\ spectra were normalized as explained in Sect.~\ref{results}. 
		Note that the scales of these diagrams are arbitrary and not quantitatively comparable. However qualitatively, the color coding in both of them is set such  that the gradual change of the color from "cold" violet  to "warmer" colors such as yellow and red corresponds to an increase of the absorption in the	sample.
		AFM denotes the antiferromagnetically ordered phase, CR marks the critical region of the
		enhanced spin fluctuations in the vicinity of the AFM phase, CPM stands for correlated paramagnet, and g-CPM denotes the gapped region of a
		depleted density of spin excitations in the correlated paramagnet. Open squares in {\bf(a)} are the estimates of the field-induced spin gap in
		temperature units from Ref.~\cite{Hentrich2017}. In {\bf(b)} the region at higher frequencies bordered by a white-dashed line denotes the broadened
		resonance modes still visible at $T = 10$\,K [cf. top curve in Fig.~\ref{fig:spectra_Fdep_Tdep}(b)], the gray dashed line sketches the area to which
		the resonance modes at $T < T_{\rm N}$ are confined with the minimum spin-wave energy gap at $\mu_0H = 7$\,T of $\Delta_{\rm sw}^{\vec{q}=0} =
		250$\,GHz (see Fig.~S4 in SM \cite{Suppl} and Refs.~\cite{Wang2017,Little2017,Ponomaryov2017}), and the horizontal white-dashed line marks the minimum energy of the
		excitation continuum probed by MWA. (for details see the text)}
	\label{fig:phase_diagram}
\end{figure*}
%

\section{Discussion \label{discus}}

Given that the MWA technique probes the field dependence of the imaginary part of the dynamic spin susceptibility $\chi^{\prime\prime}(\omega,\vec{q}= 0)$ \cite{KuboTomita1954},
a strong MWA peak developing below $T_{\rm N}$ around the critical field $\mu_0H_{\rm c} \sim 7$\,T can be related to the enhanced density of the
spin fluctuations in \rucl\ at the vicinity of the field-induced phase transition from the AFM-ordered to a non-ordered state at higher fields. At $T
> T_{\rm N}$, the decrease of the MWA at higher fields ([Figs.~\ref{fig:spectra_Fdep_Tdep}(b),(c)] implies thus a depletion of the spin excitations
probed by microwaves, i.e., a field-induced gapped behavior of $\chi^{\prime\prime}(\omega)$ which is visible in the \SD$(H)$ spectra up to
$T\sim 20 - 25$\,K.

The MWA data can be used to construct a schematic phase diagram of \rucl\ which is drawn by lines in Fig.~\ref{fig:phase_diagram} together with the
representative color-coded maps of the \SD$(H)$ spectra.  Fig.~\ref{fig:phase_diagram}(a) depicts the $T - H$ parameter space where the maps are
plotted for a representative frequency $\nu = 273$\,GHz (see also Fig.~S5 in SM \cite{Suppl}), and Fig.~\ref{fig:phase_diagram}(b) presents the $\nu - H$
cut of the MWA spectra at a temperature $T = 10$\,K $>T_{\rm N} = 8$\,K.
We note that, with the present setup, the absorption cannot be measured on an absolute scale. Also a comparison of their values measured at different
$T$ and $\nu$ is prone to a number of instrumental uncertainties. Therefore, to facilitate the pictorial presentation of the data, we plot in
Fig.~\ref{fig:phase_diagram} the normalized spectra, as explained in Sect.~\ref{results}, so that the maps $S_{\rm d}^{\rm norm}(T,H,\nu)$ in panels
(a) and (b) reflect field-induced \emph{changes} in the temperature and frequency dependence of MWA.

Keeping in mind the presence of the N\'eel transition at $T_{\rm N} = 8$\,K at $0$\,T and the sizeable spin gap $\sim 40 - 50$\,K
\cite{Baek2017,Hentrich2017} existing at $15.5$\,T, four distinct regimes enclosed by dotted lines (AFM, CR, CPM and g-CPM) with characteristic MWA
properties can be identified. Given that the probing frequencies $\nu = 80 - 360$\,GHz ($0.3 - 1.5$\,meV) considered for the phase diagram in
Fig.~\ref{fig:phase_diagram} are smaller than the spin-wave energies probed by INS at $\vec{q} = 0$ (2.7\, meV at 0\,T \cite{Banerjee2017a} and
3.4\,meV at 8\,T \cite{Banerjee2017b}), the observed non-resonant effects should be related to the coupling of microwaves to spin excitations other
than magnons. Such a field-dependent MWA below the spin-wave gap energy, which, as can be seen in Fig.~\ref{fig:phase_diagram}(b), is most
importantly present also above $T_{\rm N}$ is unexpected for a conventional antiferromagnet. The non-resonant character of the MWA suggests a
continuous energy spectrum of the probed excitations, typical for spin fractionalization \cite{fractionalizationScenario}. The vanishing of the field-dependent microwave response
below $\sim 100$\,GHz implies a characteristic minimum energy of $\sim 0.4$\,meV (100\,GHz) for this continuum [Fig.~\ref{fig:phase_diagram}(b)].
Interestingly, a remarkably similar drop-off in the optical conductivity $ \sigma_1(\omega)$ below $\sim 1$\,meV (242\,GHz), well defined even at $k_{\rm B}T \gg 1$\,meV was reported in Ref.~\cite{Little2017}. 
Such similarity supports the conjecture made there that  $ \sigma_1(\omega)$ at energies well below the range of expected optical transitions is related to the spin degree of freedom in \rucl.
Remarkably, a recent theory on the electromagnetic absorption of materials described by the Kitaev model predicts that -- similarly to the previously discussed special case of some Mott insulators (see, e.g., Refs.~\cite{Bulaevskii08,Potter13}) --  the fractionalized magnetic excitations may respond to the external {\it ac} electric field  as well \cite{Bolens18}. Within this exotic scenario, the MWA continuum observed in our experiments may manifest a coupling of both electrical and magnetic components of the sub-THz electromagnetic radiation to the fractionalized spin excitations in \rucl\ that develop a field induced spin gap. 

We note that the MWA continuum is unlikely to be explained by a coupling of one-magnon and two-magnon states in the ordered phase.
	First, the field-dependent MWA is observed also at temperatures significantly larger than $T_{\rm N}$ and, second, at energies below
	the spin-wave gap at wavevector $\vec{q} = 0$ (i.e. at the $\Gamma$-point of the Brillouin zone). Second-order effects which might yield sub-gap spin
	excitations at the $\Gamma$-point, e.g., due to the softening of the spin waves at the AFM-ordering wavevector at the $M$-point, appear unlikely
	as they would result in small-intensity features only. This conclusion is consistent with recent theory works \cite{Winter2017NatComm,Winter2017b}.

At low temperatures
the detected $S_{\rm d}$ signal exhibits an enhanced absorption of the microwaves by the sample by approaching the critical field $\mu_0H_{\rm c}\sim
7$\,T for the suppression of the AFM order. This evidences a boosted density of spin fluctuations in the critical region (CR) around this field where
the spin-wave gap $\Delta_{\rm sw}^{\vec{q}=0}$ reduces to $\sim 1$\,meV [Fig.~\ref{fig:phase_diagram}(b)]. On this strong background  it is difficult to discern a weaker response due to the interaction of microwaves with the continuum [see Fig.~S4(b) in SM \cite{Suppl}].
In the "high"-temperature regime at low
fields the interaction of microwaves with the continuum of spin excitations does not reveal a measurable field dependence, possibly due to the thermal broadening of the gap, while the continuum itself, according to INS \cite{Banerjee2017a}, still persists up to much higher energies and temperatures.
This part of the $T - H$
parameter space is denoted as correlated  paramagnet (CPM) in Fig.~\ref{fig:phase_diagram}. 
At higher fields the MWA reduces with increasing field suggesting a depletion of the density of fractionalized excitations. This region is denoted as a gapped phase of CPM (g-CPM). 
The inflection point in the MWA signal [Fig.~\ref{fig:spectra_Fdep_Tdep}(b)], which denotes the onset of the field dependence of the microwave absorption with increasing the field strength above $\sim 7$\,T, approximately demarcates in the phase diagram in Fig.~\ref{fig:phase_diagram}(a) the CPM regime at smaller fields  and the g-CPM phase at larger fields. 

 A precise quantitative determination of the gap size from MWA measurements alone is difficult because of the combined dependence on frequency and temperature. Despite this difficulty, on the more qualitative level, our data directly show that the energy scale of the field-induced gap corresponds to the temperature range where the depletion of the microwave absorption sets in. It approximately follows the dotted line in Fig.~\ref{fig:phase_diagram}(a) which separates the CPM and g-CPM regimes in $\alpha$-RuCl$_3$. This is consistent with the energy scale of the field-induced spin gap observed in the NMR relaxation rates \cite{Baek2017,Jansa2018}  and thermal conductivity \cite{Hentrich2017} measurements [open squares in Fig.~\ref{fig:phase_diagram}(a)], whereas specific heat data reveal somewhat smaller gap values \cite{Sears2017,Wolter2017}. Note that alternatively, a power law behavior of the NMR relaxation rate  was observed in Ref.~\cite{Zheng2017}.

\section{Conclusions \label{conclud}}

In summary, our field-dependent MWA study reveals a highly unusual excitation continuum in \rucl\  at the wavevector $\vec{q} = 0$ at low
energies between 0.4 and $1.2$\,meV, i.e., \emph{below} the energy of the lowest reported $\vec{q} = 0$ magnon-like mode
[Fig.~\ref{fig:phase_diagram}(b)] and extending to temperatures $T \gg T_{\rm N}$ in a broad field region up to 16\,T
[Fig.~\ref{fig:phase_diagram}(a)]. The continuum seen by MWA gets progressively gapped above $H_{\rm c}$, in line with observations of a
field-induced spin gap in Refs.~\cite{Baek2017,Sears2017,Wolter2017,Hentrich2017} which value of $~40 - 50$\,K at 15\,T is significantly larger than
the energies probed by MWA. This is  compliant with the MWA caused by a continuum below the lowest sharp mode and persisting at elevated
temperatures, a feature inconsistent with conventional magnon excitations. 
The lower bound of the continuum appears to be
quite small, of the order $\sim 100$\,GHz (0.4\,meV).
The continuum suggests a natural explanation in terms of fractionalized excitations. Given that a generic Kitaev spin liquid displays an excitation
continuum down to lowest energies \cite{Balents2016}, possibly peaked at an energy corresponding to a fraction of the Kitaev coupling
\cite{Knolle2014a}, we consider it likely that the experimental continuum represents genuine spin-liquid physics. In a fractionalization scenario, the
sharper magnon-like modes observed at elevated energy may then be interpreted as bound-state magnetic excitations \cite{Ponomaryov2017}. We emphasize
that the gapped phase above $H_{\rm c}\sim 7$\,T depicted in Fig.~\ref{fig:phase_diagram} as g-CPM is very unconventional in the sense that
the gap seen in many physical properties \cite{Baek2017,Sears2017,Wolter2017,Hentrich2017}  is not a usual spin-wave gap typically observed in
antiferromagnets  or in fully polarized states of SU(2)-symmetric magnets above a saturation field, but a gap in a continuum of excitations.
Our MWA study reveals that these broad in energy excitations are of $\vec{q} = 0$ nature which offers new insights for the interpretation of other
experimental results. Account for other interactions beyond the Kitaev model may bring additional complexity in the spin dynamics \cite{Winter2017b}.
Thus, developing a detailed theory of microwave absorption for the Kitaev spin liquid and its descendants is a key task for the
future.\\

\section*{Acknowledgments}

The authors acknowledge valuable discussions with J. van den Brink, R. Moessner, J. Knolle, I. Rousochatzakis, L. Janssen and R. Valenti. This
work has been supported by the Deutsche Forschungsgemeinschaft (DFG) via SFB 1143 and Grant No. KA 1694/8-1.\\

\appendix

\bibliography{literature_alpha-RuCl3}

\end{document}